\documentclass[11pt]{article}

\usepackage[]{acl}

\usepackage{times}
\usepackage{latexsym}

\usepackage[T1]{fontenc}

\usepackage[utf8]{inputenc}

\usepackage{microtype}

\usepackage{inconsolata}

\usepackage{graphicx}
\usepackage{subcaption}  
\usepackage{booktabs}
\usepackage{makecell}
\usepackage{xcolor}
\usepackage{tabularx}
\usepackage{listings}
\usepackage{longtable}
\usepackage{comment}

\newcommand{\catsquare}[1]{\textcolor{#1}{\rule{1.2mm}{1.2mm}}}

\definecolor{DeliberatePrioritisation}{HTML}{0072B2}
\definecolor{SelfDiagnosedFocus}{HTML}{56B4E9}
\definecolor{ThinkingStyle}{HTML}{009E73}
\definecolor{EmotionalDriver}{HTML}{D55E00}
\definecolor{InteractionAttribution}{HTML}{CC79A7}
\definecolor{DecisionNature}{HTML}{E69F00}
\definecolor{NoAccessibleReason}{HTML}{999999}
\definecolor{PatternDisputed}{HTML}{555555}
\definecolor{ContentLevelResponse}{HTML}{F0E442}

\title{Why This and Not That? A Collaborative Reflection Approach for Understanding Thought Coverage in Decision Making Support Dialog\thanks{Accepted to Findings of the Association for Computational
Linguistics: EMNLP 2026.}}

\author{
  Morita Tarvirdians \quad Hayley Hung \quad Catharine Oertel \\
  TU Delft, Delft, The Netherlands \\
  \texttt{M.Tarvirdians@tudelft.nl} \quad
  \texttt{h.hung@tudelft.nl} \quad
  \texttt{c.r.m.m.oertel@tudelft.nl}
}

\begin{document}
\maketitle

\begin{abstract}
Conversational agents that support reflection for decision-making often rely on adaptive dialog policies that map observed user behavior to actions such as probing, deepening, or redirecting. Yet the same pattern can reflect a range of different reasons such as deliberate prioritisation or limited self-access. By modeling the observable pattern rather than the user’s reason for it, current policies risk premature assumptions about the user state and inappropriate next actions.
To address this gap, we introduce a human-centered method for surfacing this hidden inference step. In a user study with 62 users and 232 collaborative moments, we pause a reflection-support agent when it would normally redirect the conversation, surface its observation, and ask users to interpret the pattern and decide how to proceed. We derive a taxonomy of nine interpretation categories and show that similar reflective states can call for substantially different follow-up actions.
Our findings challenge the assumption that adaptive dialog policies can rely on observable behavior alone, and suggest how user-provided interpretations can inform more appropriate conversational actions.

\end{abstract}

\section{Introduction}
Conversational agents are increasingly used to support people in thinking through complex situations, from health coaching~\citep{kocielnik2018reflection} to personal journaling~\citep{kim2024mindfuldiary} and decision-making~\citep{park2026choicemates}. 
These agents do more than generate contextually appropriate responses, they make \textit{strategic decisions} about the trajectory of the conversation. At each turn, the agent decides which topic to explore, whether to probe deeper or move on, and how to frame the next question. These decisions determine which aspects of a person's situation receive attention and which do not.

\begin{figure}[t]
  \centering
  \includegraphics[width=\columnwidth]{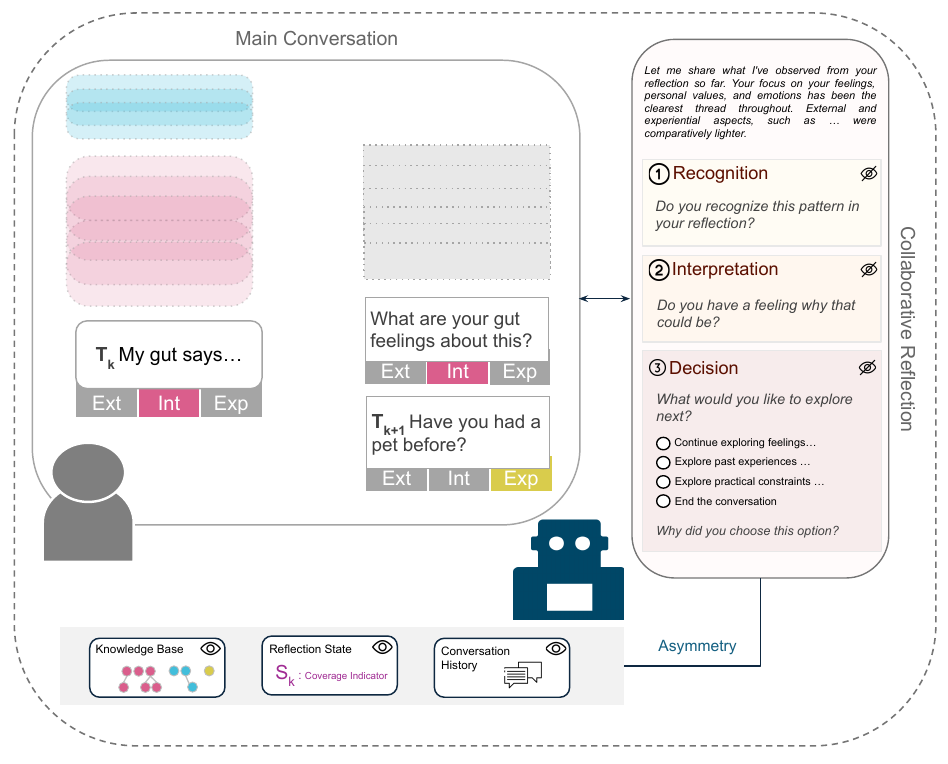}
  \caption{An illustration of a strategic decision moment in which the agent, based on observable signals (i.e., knowledge base, coverage indicator $S_k$, and conversational history after turn $T_k$), detects a coverage asymmetry across thought categories. Instead of silently acting on it, it triggers a side panel to surface its observation and collaboratively manage the decision, eliciting information not observable from dialog alone, such as whether the user recognizes the pattern, how they interpret it, and what direction they would choose for the next dialog action and why. The panel then closes, and the main conversation resumes at turn $T_{k+1}$, where the agent continues the dialog in the user-selected direction.}
  \label{fig:intro}
\end{figure}

Agents that manage dialog strategy face a fundamental tension, they must interpret the user's behavior to decide what to do next, but the reasons underlying these observable patterns are often not directly accessible to the agent.
For example, a support agent that observes sustained focus on a single topic may interpret this as fixation and attempt to redirect the conversation elsewhere (e.g., in~\citep{reflectimate}). However, the same observable pattern may reflect avoidance of an uncomfortable topic, genuine relevance, or engagement with an issue that the user has not yet resolved.
At the same time, agents that simply follow the user's stated preferences risk reinforcing blind spots; sycophantic behavior in LLMs reduces users' willingness to consider alternative perspectives and inflates their confidence in existing beliefs~\citep{sharma2024sycophancy, cheng2026sycophancy}. 
The gap between the behavioral pattern an agent observes and the user's actual reason for that pattern, what we term the \textit{interpretive gap}, remains unstudied.
 
This gap is relevant for any dialog system that acts on inferred user state. Systems that select coaching strategies based on dialog state~\citep{jorke2025gptcoach}, adapt questioning depth based on user responses~\citep{kim2024mindfuldiary}, or learn tailored strategies for diverse users~\citep{zhang2024strength, zhao2025dream} all interpret user behavior as part of their decision process. 
These systems were not designed to address the interpretive gap but the gap is present whenever an agent acts on an observable pattern whose underlying cause is ambiguous. Without understanding the space of possible reasons behind such patterns, dialog strategy operates on untested assumptions.
The challenge with interpreting a user's behavior accurately is that the set of possible reasons can be indeterminate due to ambiguity in the context of the situation. To account for ambiguity, we argue that an explicit dialog strategy is necessary to enable user and AI to collaboratively hone in on and disambiguate observed patterns of behavior in the user. 
 
In this paper, we build on ReflectiMate~\citep{reflectimate}, a reflection-support agent for decision making with a transparent and traceable policy. We make the agent’s strategic decision moments \textit{collaborative} --- specifically, the moments where the agent would otherwise silently decide which aspect of the decision should be explored next.  Rather than acting on its interpretation silently, the agent pauses and presents its observation to the user and gathers information on its recognizability and relevance to the user. Specifically, the user indicates whether they recognize the observed pattern, provides their interpretation of why it exists (their consciously aware reasons), and decides what should happen next in the dialog. 
Using data collected from 62 participants and 232 collaborative moments across diverse decision topics, we address the following research questions:

\begin{enumerate}
    \item What interpretations do users provide when the agent surfaces its observations at strategic decision points, and what taxonomy of reasons emerges?
    \item How do users' decisions align with or diverge from the agent's defaults, and what factors account for the differences?
\end{enumerate}

\section{Related Work}
\subsection{Dialog Strategy from Observable User Signals}

Conversational agents that support coaching, reflection, or decision-making must continuously decide how to steer the conversation. These strategic decisions depend on inferring the user's state from observable patterns in the dialog. However, the same observable pattern can arise from fundamentally different user states, and current systems mainly do not account for this ambiguity.

This manifests differently across systems. GPTCoach~\citep{jorke2025gptcoach} selects motivational interviewing strategies based on dialog state; when a user shows low motivation, the agent interprets this uniformly and applies an affirmation or open question. MindfulDiary~\citep{kim2024mindfuldiary} adapts follow-up questions based on what the user has shared; when a journal entry is brief, the agent interprets this as an opportunity to probe deeper. FutureMe~\citep{dechant2025future} generates contextual follow-ups to deepen reflective writing, interpreting shallow responses as warranting elaboration. ReflectiMate~\citep{reflectimate} monitors coverage across thought
categories and redirects attention when it detects asymmetry,
interpreting high coverage in one category as fixation.

The mapping from observable pattern to appropriate action is ambiguous, and the agent resolves this ambiguity with an implicit assumption rather than collaboratively with the user. How often these assumptions are wrong and what users' actual reasons are has not been empirically investigated.

\subsection{Between Deference and Override}

The problem of aligning agent behavior with user needs has been studied from two ends of a spectrum. Agents that defer to user preferences risk reinforcing blind spots: sycophantic AI reduces willingness to consider alternatives and inflates confidence in existing beliefs~\citep{sharma2024sycophancy, cheng2026sycophancy}. Agents that act on their own objectives risk ignoring user diversity: one-size-fits-all strategies fail when users have different characteristics~\citep{zhang2024strength}, and generating strategically aligned responses remains challenging when the required strategy depends on the client's unobservable state~\citep{basar2025, sun2025rethinking}. 

Between these two ends, a small body of work explores collaborative approaches. 
Collaborative management of the dialog has a long history. Plan-based models describe how participants interrupt an ongoing task to engage in a distinct \textit{subdialogue}, negotiating a proposal or resolving a disagreement before returning to the task at hand~\citep{subd2, subd1}.
Our approach shares this move. It differs in what is exchanged. 
Earlier work negotiated beliefs about a shared plan, whereas we elicit the user's reason for a pattern in their own reflective behavior, which the agent has no plan-based route to infer.

Among recent systems, the closest to ours is Deliberative AI~\citep{ma2025towards}, which proposes a process where humans and AI deliberate together to resolve conflicting perspectives. However, their deliberation concerns the task-level decision itself (what to decide), whereas our work focuses on the agent's \textit{strategic} decisions within the dialog (how to steer the conversation). We study what happens when the agent opens up the moments where it would normally act silently, producing empirical data about the gap between agent interpretations and user reasons.

\subsection{User Modeling and the Interpretive Gap}
Acting on an unobservable user state is a long-standing problem. Plan
recognition infers the user's goal from observed actions~\citep{bayesian},
and POMDP-based dialog systems extend this to action selection,
choosing what to say next under a distribution over user
states~\citep{pomdp1, pomdp}.
Recent work has modeled user emotions to improve dialog responses~\citep{lin2023emous}, learned user preferences across sessions~\citep{mehri2026multisessioncollab}, and used reinforcement learning with world models that represent latent user states such as sentiment and belief~\citep{zhao2025dream}. These approaches go beyond simple behavioral rules, they build rich representations of user state and use them to select actions.

However, even rich user models capture \textit{what} the user is doing or saying, not \textit{why}. A user who disengages from a topic because it is irrelevant needs a different response than one who disengages because the topic is emotionally uncomfortable, yet both may produce the same latent representation. The ``why'' constitutes an interpretive layer that current models lack: the mapping from observable state to the user's reason for being in that state. 

Our work addresses this gap with a collaborative approach at strategic decision moments in dialog, where the agent surfaces its observations and users provide interpretations. Rather than positing a state space and inferring over it, we characterize empirically how users interpret agent-observed patterns and select subsequent dialog actions.

\section{Methodology}
We describe our method in three subsections describing the reflection-support agent that provides the state-based policy, the determination of collaborative reflection moments that expose and query the policy’s category-switching decisions, and the user study through which we collect users’ interpretations and follow-up choices.
\subsection{Baseline Reflection Support Agent}
We build on ReflectiMate~\cite{reflectimate}, a reflection-support agent with a transparent dialog policy. 
The agent maintains a structured knowledge base (KB) of the user's thoughts, organized into three categories grounded in the decision-making literature: \textit{internal} (feelings, personal values, emotions), \textit{external} (practical constraints, other people's opinions, real-world conditions), and \textit{experiential} (past experiences, lessons from others).

The KB is constructed from the user's preliminary (unaided) reflection and updated after each conversational turn during assisted reflection phase. 
Each user utterance is parsed into one or more \textit{thoughts}, each classified into one of the three categories based on content. The agent tracks two dimensions per category $k \in \{internal, external, experiential\}$:

\begin{itemize}
    \item \textbf{Breadth} $B_k$: the number of distinct thoughts in category $k$.
    \item \textbf{Depth} $D_k$: the total number of follow-up expansions across thoughts in category $k$.
\end{itemize}
\noindent The agent computes a cumulative \textit{coverage indicator} for each category:
\begin{equation}
\label{eq:indicator}
    S_k = B_k + D_k
\end{equation}

\noindent where $S_k$ is a heuristic measure that increases both with the number of thoughts in a category (breadth) and with the number of elaborations attached to them (depth), thus reflecting the overall focus of reflection within that category. It is updated after each turn and guides the agent’s decisions.

A key objective of the ReflectiMate agent is to reduce reflection asymmetry across categories to support more comprehensive decision-making, consistent with prior work highlighting the value of integrating affective, cognitive, and experiential perspectives~\citep{dotlich2006}.

Coverage asymmetry between categories is an observable signal for the agent, computed from the KB  and the coverage indicator $S_k$. However, the underlying reason for this pattern is not observable. The agent therefore applies a uniform interpretation, assuming that higher-scoring categories are relatively overexplored and lower-scoring categories are underexplored.
Based on this interpretation, the agent selects the category $\arg\min_k S_k$ and redirects the conversation accordingly, without consulting the user about the choice of direction.
This is a strategic decision point, because the agent determines which aspect of the decision has received sufficient attention and which aspect should be further explored.

\subsection{Collaborative Reflection Moments}
To investigate how users interpret the reflection patterns observed by the agent (RQ1) and how these interpretations influence conversational direction choices (RQ2), we modified the ReflectiMate interaction pipeline by introducing \textit{collaborative reflection moments}. These moments occur whenever the agent would normally make a silent strategic decision, i.e. redirecting the conversation toward another decision aspect.

At each collaborative reflection moment, the agent externalizes the reflective pattern it observes, asks the participant to interpret the pattern, and then asks how they would like to proceed. Each moment therefore produces: (1) the observable reflection state available to the agent, (2) the participant’s interpretation of that state, and (3) the participant’s preferred conversational direction. These moments constitute the primary unit of analysis for both RQ1 and RQ2.

Collaborative reflection moments can occur at the beginning of the assisted reflection phase, if asymmetry is already present in the user’s preliminary reflection, or during the conversation when the agent’s policy would normally switch categories.

Depending on the KB state, the agent presents one of two types of observations:
\begin{itemize}
    \item \textbf{Simple Asymmetry:} one category has substantially higher coverage than the others.
    \item \textbf{Cross-category Mismatch:} despite prompting reflection within one category, the user’s responses continue to predominantly produce thoughts from another category.
\end{itemize}

\noindent The observation text adapts to the KB state at each strategic decision point.

Following the observation, participants complete a three-step protocol:

\paragraph{Step 1 --- Recognition.}
\textit{``Do you recognize this pattern in your reflection?''}
If participants disagreed with the observation, they were asked to describe the pattern they perceived instead.

\paragraph{Step 2 --- Interpretation.}
\textit{``Do you have a feeling for why that could be?''}
Participants provided an open-text explanation of the observed pattern.

\paragraph{Step 3 --- Decision.}
\textit{``What would you like to explore next?''}
Participants selected the next conversational direction from dynamically generated options corresponding to the available thought categories and optionally explained their choice.

The decision options were generated using descriptive labels rather than
category names (e.g., ``Explore your feelings, personal values, and
emotions'' instead of ``Explore internal factors''). Option order was randomized except for ``End the conversation,'' which was always presented last.
Importantly, the agent never revealed its own default interpretation or preferred conversational direction.
The protocol was refined through expert review and a pilot study with five participants using cognitive interviewing~\cite{willis2004cognitive} to identify ambiguities and improve question clarity.

Each collaborative moment produced a logged data tuple consisting of: (1) the KB state at trigger time ($S_k$, $B_k$, $D_k$ for each category), (2) the agent’s default decision (logged but hidden from the participant), (3) the observation text, and (4) the participant’s recognition response, interpretation, decision, and rationale.

\section{Experiment}
\paragraph{User Study Procedure.}
A user study was conducted on a web-based platform. After reviewing and agreeing to an informed consent form outlining the study purpose, data usage, and participant rights, participants completed a pre-study questionnaire collecting demographic information as well as the short form of the Self-Reflection and Insight Scale (SRIS)~\cite{SRIS-short}.
Participants then selected a decision topic and entered the “unaided reflection” phase, during which they freely wrote down their thoughts without any agent support. Once they indicated that they had no further thoughts to add, they proceeded to an “assisted reflection” phase, in which they interacted with the conversational agent. This interaction continued until participants chose to end the session.

\paragraph{Participants.}
We recruited 62 participants via the crowd-sourcing platform Prolific\footnote{\url{https://www.prolific.com}}.
Inclusion criteria were fluency in English and currently being in the process of making one of the big life decisions~\cite{camilleri2023investigation} listed in Appendix~\ref{app:decisionTopics}. 

The sample was diverse in terms of age (M = 34.27, SD = 8.66, median = 30, range 21–60), gender (30 female, 32 male), education level (ranging from high school to doctoral degrees), self-insight scores (M = 26.55, SD = 6.22, median = 26), and selected decision topics.

\subsection{Interpretation Taxonomy Construction}

For RQ1, the unit of analysis was each participant’s open-text interpretation response produced at a collaborative decision moment.

We developed the interpretation taxonomy using an inductive qualitative coding approach~\citep{thomas2006general}, deriving categories iteratively from participant responses. All responses were manually coded by the primary researcher. To encourage diversity during early codebook development, coding order followed maximum variation sampling~\citep{patton2014qualitative}, a strategy that deliberately selects cases expected to differ on a theoretically relevant dimension. Here, we conditioned on SRIS-S insight scores because self-insight may shape how users interpret and explain their reflective patterns, selecting participants from the high and low ends of the distribution.

For each response, the coder reviewed the full interaction context (agent observation, recognition response, and interpretation), assigned a provisional code, and recorded a brief rationale. After each batch of six participants, codes were consolidated into candidate categories with definitions and examples and iteratively refined through comparison with new data. Following~\citet{hennink2017code}, saturation was reached when six consecutive participants introduced no new categories.

\paragraph{Inter-rater reliability.}
We evaluated coding reliability using local LLMs as independent coders, following work on using LLMs as qualitative coders~\citep{than2025updating}, and recent work on LLM-assisted
reliability assessment~\citep{jain2025multi}.
We used models from four families (Qwen 2.5~\citep{qwen2.5}, Mistral~\citep{jiang2023mistral}, Phi-4~\citep{abdin2024phi}, and Gemma 2~\citep{gemma2}) to reduce shared-bias effects~\cite{than2025updating}. Each model independently coded all responses using the final codebook without access to researcher labels, and all runs were executed locally via Ollama to ensure data privacy.

Agreement between the human coder and the individual models ranged from $\kappa = 0.42$ to $\kappa = 0.50$, with the highest inter-model agreement at $\kappa = 0.58$ (Appendix~\ref{app:irr}).

\subsection{Reflection Trajectory Analysis}

To investigate alignment between user-selected and agent-selected reflection trajectories (RQ2), we compared the participant’s chosen conversational direction against the agent’s logged default decision (according to $S_k$) at each collaborative moment.

We report overall alignment rates as well as alignment rates conditioned on interpretation category. To examine whether interpretation categories mediate the relationship between observable reflection states and trajectory choices, we additionally analyze the distribution of user-selected directions within each dominant reflection state.

Finally, we characterize the nature of trajectory divergence by classifying each user choice as either: (1) moving toward an underexplored reflection category, or (2) remaining within the dominant category. We report these distributions overall and across interpretation categories.

\section{Results}
\begin{table*}

\footnotesize
\centering

\begin{tabularx}{\textwidth}{@{}l X X r@{}}
\toprule
\textbf{Category} & \textbf{Condensed Description} & \textbf{Example} & \textbf{Distribution}\\
\midrule

\catsquare{DeliberatePrioritisation} Deliberate Prioritisation &
Conscious judgment about what to focus on &
\makecell[l]{"My personal experience is \\ most important"} &
$n$=64 (27.6\%) \\

\midrule

\catsquare{SelfDiagnosedFocus} Self-Diagnosed Focus &
Diagnoses a personal psychological tendency &
\makecell[l]{"I think when I feel negative emotions \\ I try very hard to rationalise them \\ rather than confront them"} &
$n$=46 (19.8\%) \\

\midrule

\catsquare{EmotionalDriver} Emotional Driver &
Emotion or need named as reason &
\makecell[l]{"Guilt"} &
$n$=32 (13.8\%) \\

\midrule

\catsquare{DecisionNature} Decision Nature &
Nature/circumstances of the decision &
\makecell[l]{"Out of necessity, to pay bills"} &
$n$=26 (11.2\%) \\

\midrule

\catsquare{ThinkingStyle} Thinking Style &
Habitual cognitive approach &
\makecell[l]{"That's just the way my brain works"} &
$n$=20 (8.6\%) \\

\midrule

\catsquare{NoAccessibleReason} No Accessible Reason &
Cannot articulate why the pattern exists &
\makecell[l]{"I really don't know why"} &
$n$=16 (6.9\%) \\

\midrule

\catsquare{InteractionAttribution} Interaction Attribution &
Reason attributed to structure or state of the interaction itself &
\makecell[l]{"I don't feel the conversation \\ has progressed to this stage yet"} &
$n$=15 (6.5\%) \\

\midrule

\catsquare{PatternDisputed} Pattern Disputed &
Rejects the agent's characterization &
\makecell[l]{"I was focusing on experiences,\\ not feelings"} &
$n$=9 (3.9\%) \\

\midrule

\catsquare{ContentLevelResponse} Content-Level Response &
No meta-cognitive explanation; response remains at surface content level &
\makecell[l]{"I know deep down that \\it's the right long term decision"} &
$n$=4 (1.7\%) \\

\bottomrule
\end{tabularx}
\caption{Taxonomy of interpretation categories ($N$=232), ordered by frequency, with condensed definitions, illustrative examples, and distribution of occurrences.}
\label{tab:rq1}
\end{table*}

\subsection{Taxonomy of User Interpretations (RQ1)}

From 232 collaborative reflection moments, we derived a taxonomy of 9 interpretation categories (Table~\ref{tab:rq1}; full taxonomy in Appendix~\ref{app:taxanomony}). Inductive coding stabilized after 18 participants, with no new categories emerging from participants 19--62.

A central finding was that similar observable reflection states often corresponded to substantially different user interpretations. When grouping moments by shared observable KB states, 28 groups contained at least two different interpretation categories, and 18 groups contained three or more categories. This indicates that the interpretive gap was pervasive rather than limited to isolated cases.

The most frequent category was \textit{Deliberate Prioritisation} ($n=64$, 27.6\%), where users described their reflective focus as an intentional judgment about which aspect of the decision deserved greater attention. \textit{Self-Diagnosed Focus} ($n=46$, 19.8\%) captured cases in which users analytically explained the source of their reflective pattern, while \textit{Emotional Driver} ($n=32$, 13.8\%) captured cases where users primarily attributed the pattern to an emotion without further analysis. Together, these three categories accounted for 61.2\% of all collaborative moments.

To further examine interpretive diversity under similar observable conditions, we analyzed moments sharing the same dominant category, observation type, and comparable coverage-score gaps. The most diverse group --- characterized by a dominant internal category and a score gap of approximately 10 --- contained 19 moments spanning 8 of the 9 taxonomy categories. Interpretations ranged from \textit{“it’s nice to talk about my feelings”} (\textit{Emotional Driver}) to \textit{“just the way I am”} (\textit{Thinking Style}) to \textit{“I really don’t know”} (\textit{No Accessible Reason}).

Despite this diversity, the agent’s policy would have treated all 19 moments identically, as they corresponded to the same observable reflection state from the agent’s perspective.

\begin{figure*}[t]
  \centering
  \begin{subfigure}[t]{0.44\textwidth}
    \includegraphics[width=\linewidth]{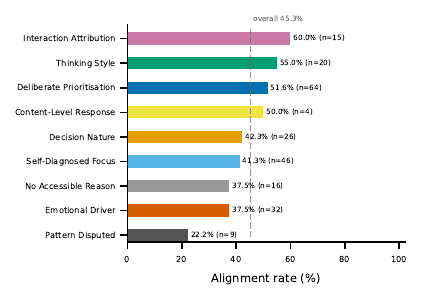}
    \caption{Alignment rate by taxonomy category. Dashed line = overall rate (45.3\%).}
    \label{fig:rq2a}
  \end{subfigure}
  \hfill
  \begin{subfigure}[t]{0.54\textwidth}
    \includegraphics[width=\linewidth]{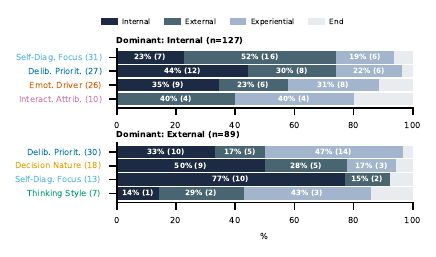}
    \caption{User choice direction by taxonomy category, split by dominant
    KB state.  Row labels give the number of moments in each category; segment labels give percentage and raw count. The four most frequent categories are shown in each panel.}
    \label{fig:rq2b}
  \end{subfigure}
  \caption{Alignment and divergence behavior across taxonomy categories.}
  \label{fig:rq2}
\end{figure*}

\subsection{Alignment and User Choices (RQ2)}

Users selected the same action as the agent’s default in 105 of 232 moments (45.3\%). In the remaining cases, users selected a different direction than the coverage-based policy.

\paragraph{Divergence does not imply resistance to exploration.}
Importantly, disagreement with the agent did not mean users stayed with what they had already explored. Among the 127 misaligned moments, 45.7\% moved
toward an underexplored dimension (albeit not the one selected by the agent), 41.7\% remained within the dominant dimension, and 12.6\%
ended the conversation. 
Across moments in which the conversation continued, 74.4\% (160/215) of user choices moved toward a dimension other than the current dominant one. The agent's observation therefore influenced \textit{whether} users broadened their exploration, while the direction of broadening varied.

\paragraph{Interpretation is associated with divergence patterns.}
Alignment rates varied across taxonomy categories, ranging from 22.2\% to 60.0\% (Figure~\ref{fig:rq2a}). This indicates that the likelihood of following the agent’s default decision depends in part on how users interpret the observed reflection pattern.

Figure~\ref{fig:rq2b} further shows that user choices differ systematically within the same observable state depending on interpretation category. For example, when internal aspects dominated reflection ($n=127$), users attributing the pattern to \textit{Self-Diagnosed Focus} most frequently shifted toward external aspects (52\%), whereas those attributing it to \textit{Deliberate Prioritisation} more often continued exploring internal aspects (44\%). When external aspects were dominant ($n=89$), users attributing the pattern to \textit{Self-Diagnosed Focus} never selected a shift toward experiential aspects, whereas other interpretation categories did.

Overall, these results suggest that the same observable reflection state can lead to different trajectory choices depending on the user’s interpretation, which is not accessible to the agent’s policy.

\subsection{Exploratory Analyses}

\paragraph{Recognition and observation type.}
Users recognized the agent’s observation in 89.2\% of collaborative moments. Recognition was universal for simple asymmetry cases, but all non-recognition cases (25 moments) occurred in cross-category mismatch observations.

Notably, lack of recognition did not prevent users from providing interpretations: 17 of 25 such cases still contained explicit explanations, including \textit{Deliberate Prioritisation} (5/25) and \textit{Self-Diagnosed Focus} (3/25). Only a minority explicitly attributed the mismatch to incorrect pattern detection.

\paragraph{Preliminary indicators of interpretation patterns.}
We explored whether observable pre-interaction signals may be associated with interpretation categories. Users with longer preliminary reflections were more likely to produce \textit{Self-Diagnosed Focus} interpretations (33.8\% in the highest tertile vs.\ 17.1\% in the lowest) and less likely to produce \textit{Emotional Driver} responses (5.2\% vs.\ 17.1\%; $\chi^2 = 24.8$, $df = 16$, $p = .073$, $V = .23$).

A similar but weaker trend was observed for SRIS insight scores: \textit{Emotional Driver} interpretations were associated with lower insight scores ($M = 23.4$) than \textit{Self-Diagnosed Focus} interpretations ($M = 26.1$, $p = .14$).

These results are exploratory and should be interpreted as hypothesis-generating. They suggest that pre-interaction signals such as reflection length and insight may be weakly associated with how users interpret reflection patterns.

\section{Discussion}
In this paper we investigated what interpretations users give when an agent surfaces the reflection patterns it has observed, and how those interpretations relate to the next dialog action they choose: whether to redirect, or stay with the current focus. 

\subsection{A Taxonomy of User Reasons Accounting for Coverage Asymmetries}
To answer \textbf{RQ1}, we examined the reasons users give when an agent makes an observed asymmetry in their reflective behavior explicit. From these responses, we derive a taxonomy of nine interpretation categories (cf. Table~\ref{tab:rq1}), showing that the same behavioral pattern can reflect structured and recurring differences in user reasons. This opens a new modeling direction for adaptive dialog policy: agents can represent likely reason categories behind observable patterns, enabling them to choose follow-up strategies that better fit the user’s situation and the task. For instance, a follow-up strategy should be different if the reason for coverage asymmetry lies in a \textit{deliberate prioritization} in contrast to an \textit{emotional driver}.

We also provide empirical evidence for the existence of the interpretation gap between the agent and the user. For example, a single observable agent state, i.e. the same pattern in the knowledge-base representation that the agent uses to make strategic decisions, was interpreted in eight different ways by users. 
Therefore, dialog systems that select dialog trajectories solely based on observable signals, such as latent representations of inferred user beliefs~\citep{zhao2025dream}, latent representations of user coverage patterns~\citep{reflectimate}, or LLMs that prematurely lock into assumptions about user intent~\citep{laban2026llms}, risk offering uniform solutions to fundamentally different underlying user states.
While \citep{zhang2024strength} argue that one-size-fits-all strategies fail for diverse users, our findings suggest that one mechanism underlying such failures is that the same behavior can arise from different reasons, such as deliberate focus, emotional states, cognitive habits, situational constraints, and properties of the interaction itself.

\subsection{{Alignment and Divergence Between User Choices and State-Based Policy Actions}}
In \textbf{RQ2}, we investigated users’ decisions regarding the reflection trajectory and the alignment between their choices and the trajectory the agent would have selected. 
A key finding was that, when users were informed of the pattern the agent observed in their reflection and asked to interpret it, 74.4\% of choices moved toward an underexplored dimension of their decision, without requiring the agent to act unilaterally or explicitly reveal its default choice.
Yet in 54.7\% of moments users chose a different direction than the agent's default. Users broadened their exploration, but the direction was informed by reasons the agent could not have inferred from its state representation. In other words, although the agent could identify potentially underexplored regions of reflection, users’ interpretations shaped how they chose to engage with those regions. 


\subsection{A Hybrid Intelligence Approach to Dialog Strategy}
Our design and findings suggest a hybrid intelligence approach to dialog strategy: the agent does not treat its interpretation of user behavior as sufficient for action, but surfaces the observed pattern and asks the user to help interpret it before the next dialog move is selected. The agent contributes structural awareness of the reflection process; the user contributes self-knowledge about the reason behind the pattern. This is not merely a way to make the agent more agreeable: surfacing the pattern can also create meta-awareness for the user, helping them recognize how their own reflection is unfolding and giving them a stronger role in steering the conversation.

This perspective differs from approaches that emphasize collaboration at the level of task outcomes, such as joint decision-making~\citep{ma2025towards}, whereas we instead propose focusing on collaboration at the level of dialog strategy.
In addition, our approach lies between two dialog strategy paradigms in prior work, systems that solely follow user preferences, the sycophantic agents~\citep{sharma2024sycophancy, cheng2026sycophancy}, and systems that follow their own interaction objectives~\citep{reflectimate}.

Together with the taxonomy, our proposed approach opens up opportunities for designing future agents that reason over multiple plausible interpretations of user states rather than collapsing them into a single assumption. Such agents could surface uncertainty when necessary and empower users in gaining further awareness in the interpretation process itself.


\section{Future Work}

The taxonomy maps the interpretive space that adaptive dialog policies would need to account for. This paper characterizes that space and does not evaluate whether acting on it improves dialog outcomes. Establishing that requires a controlled comparison against an agent that steers
silently, together with an account of what counts as improvement. 
For big life decisions, for example, decision quality cannot be established within a single reflection session, since outcomes unfold over months or years, so evaluation would rely on process-level proxies. 
This line of experimentation is an important direction for future work.

Beyond this comparison, future work could condition action selection on inferred reason types rather than on observable state alone. The observed association between signals (e.g., reflection length, insight scores) and interpretation types further suggests that partial adaptation without explicit exchange may be feasible, though this remains preliminary and requires validation on larger samples. Finally, extending the taxonomy to other support domains would further test its generality beyond the present setting.

\section{Conclusion}
We presented the first empirical characterization of the interpretive gap in support dialog, the gap between the behavioral patterns an agent observes and the reasons underlying them. By surfacing agent observations to 62 users across 232 strategic decision moments, we derive a 9-category taxonomy of user reasons, showing that identical observable states can reflect deliberate choices, emotional states, cognitive habits, situational constraints, or interactional structure. Users diverged from the agent’s default action in 54.7\% of cases, yet 74.4\% of continuing choices shifted toward an underexplored dimension, suggesting disagreement concerned direction rather than integration. These findings highlight that dialog strategy in support agents operates over inherently ambiguous signals, and that structured agent–user exchange reveals information neither side can access alone.

\section*{Limitations}
Several limitations constrain the generalizability of our findings. The taxonomy was derived from a single agent in a single domain (personal decision-making reflection), and while it captures general cognitive phenomena such as deliberate prioritisation, emotional influence, and habitual thinking, their relative distribution may differ in other contexts such as therapeutic dialog or educational tutoring.

Our double-coding approach used local LLMs rather than a second human coder. While inter-model agreement supports codebook clarity, moderate kappa values reflect the inherent difficulty of distinguishing nuanced categories such as Emotional Driver and Self-Diagnosed Focus, a boundary that also challenges automated classification.

User-reported reasons are self-reports and may reflect post-hoc rationalization rather than actual causes of behavior~\cite{nisbett1977telling}. In addition, the study protocol may have influenced responses by prompting users to recognize patterns that might not have emerged spontaneously.

\section*{Ethical Considerations}
This study was approved by the Human Research Ethics Committee of Delft University of Technology. All participants provided informed consent before participation and were compensated in accordance with Prolific's fair payment policies.  
Given the sensitivity of personal decision-making topics, we took several precautions. When participants requested opinions or advice, the agent explicitly stated it could not provide any. All language models used for response generation and qualitative coding were deployed locally to ensure that reflection data remained on secure institutional servers and was not transmitted to third-party APIs. Data were pseudonymised, and personally identifiable information was stored separately from reflection content; participants were informed of data handling and retention practices during consent.

\section*{Gen AI Declaration}
Generative AI tools were used to assist with copy-editing and coding, but the authors are fully accountable for the content and ideas.

\section*{Acknowledgments}
We thank Catholijn M. Jonker for her continued support of this line of work. We also gratefully acknowledge the support of the Delft AI Initiative and the Designing Intelligence (DI) lab.

This research was partially funded by the Hybrid Intelligence Center, a 10-year programme funded by the Dutch Ministry of Education, Culture and Science through the Netherlands Organisation for Scientific Research, \href{https://www.hybrid-intelligence-centre.nl}{Hybrid
Intelligence Centre}, grant number 024.004.022. 
It has also been partially supported by the project RISE: Relapse Intervention and Self-Management, file number KICH1.GZ03.21.002 of the research programme KIC - MISSIE Zorg in eigen
leefomgeving 2021, partly financed by the Dutch Research Council (NWO).

\bibliography{custom}

@inproceedings{sharma2024sycophancy,
  title     = {Towards Understanding Sycophancy in Language Models},
  author    = {Sharma, Mrinank and
               Tong, Meg and
               Korbak, Tomasz and
               Duvenaud, David and
               Askell, Amanda and
               Bowman, Samuel R. and
               Cheng, Newton and
               Durmus, Esin and
               Hatfield-Dodds, Zac and
               Johnston, Scott R. and
               Kravec, Shauna and
               Maxwell, Timothy and
               McCandlish, Sam and
               Ndousse, Kamal and
               Rausch, Oliver and
               Schiefer, Nicholas and
               Yan, Da and
               Zhang, Miranda and
               Perez, Ethan},
  booktitle = {International Conference on Learning Representations},
  year      = {2024},
 url = {https://proceedings.iclr.cc/paper_files/paper/2024/file/0105f7972202c1d4fb817da9f21a9663-Paper-Conference.pdf},
}

@inproceedings{lin2023emous,
  title={EmoUS: Simulating user emotions in task-oriented dialogues},
  author={Lin, Hsien-Chin and Feng, Shutong and Geishauser, Christian and Lubis, Nurul and van Niekerk, Carel and Heck, Michael and Ruppik, Benjamin and Vukovic, Renato and Gasi{\'c}, Milica},
  booktitle={Proceedings of the 46th International ACM SIGIR Conference on Research and Development in Information Retrieval},
  pages={2526--2531},
  year={2023}
}

@article{mehri2026multisessioncollab,
  title={MultiSessionCollab: Learning User Preferences with Memory to Improve Long-Term Collaboration},
  author={Mehri, Shuhaib and Kargupta, Priyanka and August, Tal and Hakkani-T{\"u}r, Dilek},
  journal={arXiv preprint arXiv:2601.02702},
  year={2026}
}

@inproceedings{zhang2024strength,
  title={Strength lies in differences! improving strategy planning for non-collaborative dialogues via diversified user simulation},
  author={Zhang, Tong and Huang, Chen and Deng, Yang and Liang, Hongru and Liu, Jia and Wen, Zujie and Lei, Wenqiang and Chua, Tat-Seng},
  booktitle={Proceedings of the 2024 Conference on Empirical Methods in Natural Language Processing},
  pages={424--444},
  year={2024}
}

@inproceedings{jorke2025gptcoach,
  title={GPTCoach: towards LLM-based physical activity coaching},
  author={J{\"o}rke, Matthew and Sapkota, Shardul and Warkenthien, Lyndsea and Vainio, Niklas and Schmiedmayer, Paul and Brunskill, Emma and Landay, James A},
  booktitle={Proceedings of the 2025 CHI conference on human factors in computing systems},
  pages={1--46},
  year={2025}
}

@article{kocielnik2018reflection,
  title={Reflection companion: a conversational system for engaging users in reflection on physical activity},
  author={Kocielnik, Rafal and Xiao, Lillian and Avrahami, Daniel and Hsieh, Gary},
  journal={Proceedings of the ACM on Interactive, Mobile, Wearable and Ubiquitous Technologies},
  volume={2},
  number={2},
  pages={1--26},
  year={2018},
  publisher={ACM New York, NY, USA}
}

@inproceedings{kim2024mindfuldiary,
  title={MindfulDiary: harnessing large language model to support psychiatric patients' journaling},
  author={Kim, Taewan and Bae, Seolyeong and Kim, Hyun Ah and Lee, Su-woo and Hong, Hwajung and Yang, Chanmo and Kim, Young-Ho},
  booktitle={Proceedings of the 2024 CHI Conference on Human Factors in Computing Systems},
  pages={1--20},
  year={2024}
}

@inproceedings{laban2026llms,
 author = {Laban, Philippe and Hayashi, Hiroaki and Zhou, Yingbo and Neville, Jennifer},
 booktitle = {International Conference on Learning Representations},
 editor = {C. Vondrick and B. Hariharan and C. Raffel and L. Pinto and D. Yang and A. Faust},
 title = {LLMs Get Lost In Multi-Turn Conversation},
 url = {https://proceedings.iclr.cc/paper_files/paper/2026/file/59f6421e64707225fdf5b28840679a07-Paper-Conference.pdf},
 year = {2026}
}

@article{cheng2026sycophancy,
author    = {Cheng, Myra and Lee, Cinoo and Khadpe, Pranav and Yu, Sunny and Han, Dyllan and Jurafsky, Dan},
title     = {Sycophantic {AI} Decreases Prosocial Intentions and Promotes Dependence},
journal   = {Science},
volume    = {391},
number    = {6792},
pages     = {eaec8352},
year      = {2026},
doi       = {10.1126/science.aec8352}
}

@inproceedings{zhao2025dream,
    title = "Dream to Chat: Model-based Reinforcement Learning on Dialogues with User Belief Modeling",
    author = "Zhao, Yue  and
      Wang, Xiaoyu  and
      Wang, Dan  and
      Jiang, Zhonglin  and
      Gu, Qingqing  and
      Chen, Teng  and
      Xi, Ningyuan  and
      Qu, Jinxian  and
      Chen, Yong  and
      Ji, Luo",
    editor = "Christodoulopoulos, Christos  and
      Chakraborty, Tanmoy  and
      Rose, Carolyn  and
      Peng, Violet",
    booktitle = "Findings of the Association for Computational Linguistics: EMNLP 2025",
    month = nov,
    year = "2025",
    address = "Suzhou, China",
    publisher = "Association for Computational Linguistics",
    url = "https://aclanthology.org/2025.findings-emnlp.256/",
    doi = "10.18653/v1/2025.findings-emnlp.256",
    pages = "4764--4781",
    ISBN = "979-8-89176-335-7"
}

@article{dechant2025future,
  title={Future Me, a prospection-based chatbot to promote mental well-being in youth: two exploratory user experience studies},
  author={Dechant, Martin and Lash, Eva and Shokr, Sarah and O'Driscoll, Ciar{\'a}n},
  journal={JMIR Formative Research},
  volume={9},
  number={1},
  pages={e74411},
  year={2025},
  publisher={JMIR Publications Inc., Toronto, Canada}
}

@inproceedings{park2026choicemates,
  title={Choicemates: Supporting unfamiliar online decision-making with multi-agent conversational interactions},
  author={Park, Jeongeon and Min, Bryan and Son, Kihoon and Song, Jean Y and Ma, Xiaojuan and Kim, Juho},
  booktitle={Proceedings of the 31st International Conference on Intelligent User Interfaces},
  pages={1526--1550},
  year={2026}
}

@inproceedings{ma2025towards,
  title={Towards human-ai deliberation: Design and evaluation of llm-empowered deliberative ai for ai-assisted decision-making},
  author={Ma, Shuai and Chen, Qiaoyi and Wang, Xinru and Zheng, Chengbo and Peng, Zhenhui and Yin, Ming and Ma, Xiaojuan},
  booktitle={Proceedings of the 2025 CHI Conference on Human Factors in Computing Systems},
  pages={1--23},
  year={2025}
}

@inproceedings{basar2025,
  title={How well can large language models reflect? A human evaluation of LLM-generated reflections for motivational interviewing dialogues},
  author={Bașar, Erkan and Sun, Xin and Hendrickx, Iris and de Wit, Jan and Bosse, Tibor and De Bruijn, Gert-Jan and Bosch, Jos A and Krahmer, Emiel},
  booktitle={Proceedings of the 31st international conference on computational linguistics},
  pages={1964--1982},
  year={2025}
}

@book{willis2004cognitive,
  title={Cognitive Interviewing: A Tool for Improving Questionnaire Design},
  author={Willis, G.B.},
  isbn={9780761928041},
  lccn={2004013649},
  url={https://books.google.nl/books?id=yg2NDzp9zPAC},
  year={2005},
  publisher={SAGE Publications}
}

@article{camilleri2023investigation,
  title={An investigation of big life decisions},
  author={Camilleri, Adrian R},
  journal={Judgment and Decision Making},
  volume={18},
  pages={e32},
  year={2023}
}

@book{patton2014qualitative,
  title={Qualitative research \& evaluation methods: Integrating theory and practice},
  author={Patton, Michael Quinn},
  year={2014},
  publisher={Sage publications}
}

@article{hennink2017code,
  title={Code saturation versus meaning saturation: how many interviews are enough?},
  author={Hennink, Monique M and Kaiser, Bonnie N and Marconi, Vincent C},
  journal={Qualitative health research},
  volume={27},
  number={4},
  pages={591--608},
  year={2017},
  publisher={Sage Publications Sage CA: Los Angeles, CA}
}

@article{than2025updating,
  title={Updating “the future of coding”: Qualitative coding with generative large language models},
  author={Than, Nga and Fan, Leanne and Law, Tina and Nelson, Laura K and McCall, Leslie},
  journal={Sociological Methods \& Research},
  volume={54},
  number={3},
  pages={849--888},
  year={2025},
  publisher={SAGE Publications Sage CA: Los Angeles, CA}
}

@inproceedings{reflectimate,
  title={Reflecti-Mate: A Conversational Agent for Adaptive Decision-Making Support Through System 1 and System 2 Thinking},
  author={Tarvirdians, Morita and Chandrasegaran, Senthil and Hung, Hayley and Jonker, Catholijn M and Oertel, Catharine},
  booktitle={Proceedings of the 34th ACM Conference on User Modeling, Adaptation and Personalization},
  pages={213--222},
  year={2026}
}

@article{nisbett1977telling,
  title={Telling more than we can know: Verbal reports on mental processes.},
  author={Nisbett, Richard E and Wilson, Timothy D},
  journal={Psychological review},
  volume={84},
  number={3},
  pages={231},
  year={1977},
  publisher={American Psychological Association}
}

@book{dotlich2006,
  title={Head, Heart and Guts: How the world's best companies develop complete leaders},
  author={Dotlich, David L and Cairo, Peter C and Rhinesmith, Stephen H},
  year={2006},
  publisher={John Wiley \& Sons}
}

@article{SRIS-short,
  title={The self-reflection and insight scale: Applying item response theory to craft an efficient short form},
  author={Silvia, Paul J},
  journal={Current Psychology},
  volume={41},
  number={12},
  pages={8635--8645},
  year={2022},
  publisher={Springer}
}

@misc{qwen2.5,
      title={Qwen2.5-Coder Technical Report}, 
      author={Binyuan Hui and Jian Yang and Zeyu Cui and Jiaxi Yang and Dayiheng Liu and Lei Zhang and Tianyu Liu and Jiajun Zhang and Bowen Yu and Keming Lu and Kai Dang and Yang Fan and Yichang Zhang and An Yang and Rui Men and Fei Huang and Bo Zheng and Yibo Miao and Shanghaoran Quan and Yunlong Feng and Xingzhang Ren and Xuancheng Ren and Jingren Zhou and Junyang Lin},
      year={2024},
      eprint={2409.12186},
      archivePrefix={arXiv},
      primaryClass={cs.CL},
      url={https://arxiv.org/abs/2409.12186}, 
}

@misc{jiang2023mistral,
      title={Mistral 7B}, 
      author={Albert Q. Jiang and Alexandre Sablayrolles and Arthur Mensch and Chris Bamford and Devendra Singh Chaplot and Diego de las Casas and Florian Bressand and Gianna Lengyel and Guillaume Lample and Lucile Saulnier and Lélio Renard Lavaud and Marie-Anne Lachaux and Pierre Stock and Teven Le Scao and Thibaut Lavril and Thomas Wang and Timothée Lacroix and William El Sayed},
      year={2023},
      eprint={2310.06825},
      archivePrefix={arXiv},
      primaryClass={cs.CL},
      url={https://arxiv.org/abs/2310.06825}, 
}

@misc{abdin2024phi,
      title={Phi-4 Technical Report}, 
      author={Marah Abdin and Jyoti Aneja and Harkirat Behl and Sébastien Bubeck and Ronen Eldan and Suriya Gunasekar and Michael Harrison and Russell J. Hewett and Mojan Javaheripi and Piero Kauffmann and James R. Lee and Yin Tat Lee and Yuanzhi Li and Weishung Liu and Caio C. T. Mendes and Anh Nguyen and Eric Price and Gustavo de Rosa and Olli Saarikivi and Adil Salim and Shital Shah and Xin Wang and Rachel Ward and Yue Wu and Dingli Yu and Cyril Zhang and Yi Zhang},
      year={2024},
      eprint={2412.08905},
      archivePrefix={arXiv},
      primaryClass={cs.CL},
      url={https://arxiv.org/abs/2412.08905}, 
}

@misc{gemma2,
      title={Gemma 2: Improving Open Language Models at a Practical Size}, 
      author={Gemma Team and Morgane Riviere and Shreya Pathak and Pier Giuseppe Sessa and Cassidy Hardin and Surya Bhupatiraju and Léonard Hussenot and Thomas Mesnard and Bobak Shahriari and Alexandre Ramé and Johan Ferret and Peter Liu and Pouya Tafti and Abe Friesen and Michelle Casbon and Sabela Ramos and Ravin Kumar and Charline Le Lan and Sammy Jerome and Anton Tsitsulin and Nino Vieillard and Piotr Stanczyk and Sertan Girgin and Nikola Momchev and Matt Hoffman and Shantanu Thakoor and Jean-Bastien Grill and Behnam Neyshabur and Olivier Bachem and Alanna Walton and Aliaksei Severyn and Alicia Parrish and Aliya Ahmad and Allen Hutchison and Alvin Abdagic and Amanda Carl and Amy Shen and Andy Brock and Andy Coenen and Anthony Laforge and Antonia Paterson and Ben Bastian and Bilal Piot and Bo Wu and Brandon Royal and Charlie Chen and Chintu Kumar and Chris Perry and Chris Welty and Christopher A. Choquette-Choo and Danila Sinopalnikov and David Weinberger and Dimple Vijaykumar and Dominika Rogozińska and Dustin Herbison and Elisa Bandy and Emma Wang and Eric Noland and Erica Moreira and Evan Senter and Evgenii Eltyshev and Francesco Visin and Gabriel Rasskin and Gary Wei and Glenn Cameron and Gus Martins and Hadi Hashemi and Hanna Klimczak-Plucińska and Harleen Batra and Harsh Dhand and Ivan Nardini and Jacinda Mein and Jack Zhou and James Svensson and Jeff Stanway and Jetha Chan and Jin Peng Zhou and Joana Carrasqueira and Joana Iljazi and Jocelyn Becker and Joe Fernandez and Joost van Amersfoort and Josh Gordon and Josh Lipschultz and Josh Newlan and Ju-yeong Ji and Kareem Mohamed and Kartikeya Badola and Kat Black and Katie Millican and Keelin McDonell and Kelvin Nguyen and Kiranbir Sodhia and Kish Greene and Lars Lowe Sjoesund and Lauren Usui and Laurent Sifre and Lena Heuermann and Leticia Lago and Lilly McNealus and Livio Baldini Soares and Logan Kilpatrick and Lucas Dixon and Luciano Martins and Machel Reid and Manvinder Singh and Mark Iverson and Martin Görner and Mat Velloso and Mateo Wirth and Matt Davidow and Matt Miller and Matthew Rahtz and Matthew Watson and Meg Risdal and Mehran Kazemi and Michael Moynihan and Ming Zhang and Minsuk Kahng and Minwoo Park and Mofi Rahman and Mohit Khatwani and Natalie Dao and Nenshad Bardoliwalla and Nesh Devanathan and Neta Dumai and Nilay Chauhan and Oscar Wahltinez and Pankil Botarda and Parker Barnes and Paul Barham and Paul Michel and Pengchong Jin and Petko Georgiev and Phil Culliton and Pradeep Kuppala and Ramona Comanescu and Ramona Merhej and Reena Jana and Reza Ardeshir Rokni and Rishabh Agarwal and Ryan Mullins and Samaneh Saadat and Sara Mc Carthy and Sarah Cogan and Sarah Perrin and Sébastien M. R. Arnold and Sebastian Krause and Shengyang Dai and Shruti Garg and Shruti Sheth and Sue Ronstrom and Susan Chan and Timothy Jordan and Ting Yu and Tom Eccles and Tom Hennigan and Tomas Kocisky and Tulsee Doshi and Vihan Jain and Vikas Yadav and Vilobh Meshram and Vishal Dharmadhikari and Warren Barkley and Wei Wei and Wenming Ye and Woohyun Han and Woosuk Kwon and Xiang Xu and Zhe Shen and Zhitao Gong and Zichuan Wei and Victor Cotruta and Phoebe Kirk and Anand Rao and Minh Giang and Ludovic Peran and Tris Warkentin and Eli Collins and Joelle Barral and Zoubin Ghahramani and Raia Hadsell and D. Sculley and Jeanine Banks and Anca Dragan and Slav Petrov and Oriol Vinyals and Jeff Dean and Demis Hassabis and Koray Kavukcuoglu and Clement Farabet and Elena Buchatskaya and Sebastian Borgeaud and Noah Fiedel and Armand Joulin and Kathleen Kenealy and Robert Dadashi and Alek Andreev},
      year={2024},
      eprint={2408.00118},
      archivePrefix={arXiv},
      primaryClass={cs.CL},
      url={https://arxiv.org/abs/2408.00118}, 
}

@article{thomas2006general,
  title={A general inductive approach for analyzing qualitative evaluation data},
  author={Thomas, David R},
  journal={American journal of evaluation},
  volume={27},
  number={2},
  pages={237--246},
  year={2006},
  publisher={Sage Publications Sage CA: Thousand Oaks, CA}
}

@article{jain2025multi,
  title={Multi-LLM Thematic Analysis with Dual Reliability Metrics: Combining Cohen's Kappa and Semantic Similarity for Qualitative Research Validation},
  author={Jain, Nilesh and Suh, Hyungil and Adeyinka, Seyi and Roseman, Leor and Allsop, Aza},
  journal={arXiv preprint arXiv:2512.20352},
  year={2025}
}

@inproceedings{sun2025rethinking,
  title={Rethinking the alignment of psychotherapy dialogue generation with motivational interviewing strategies},
  author={Sun, Xin and Tang, Xiao and El Ali, Abdallah and Li, Zhuying and Ren, Pengjie and de Wit, Jan and Pei, Jiahuan and Bosch, Jos A},
  booktitle={Proceedings of the 31st International Conference on Computational Linguistics},
  pages={1983--2002},
  year={2025}
}

@article{bayesian,
   author = {Charniak, Eugene and Goldman, Robert P},
   title = {A Bayesian model of plan recognition},
   journal = {Artificial Intelligence},
   volume = {64},
   number = {1},
   pages = {53-79},
   ISSN = {0004-3702},
   year = {1993},
   type = {Journal Article}
}

@inproceedings{subd1,
   author = {Lambert, Lynn and Carberry, Sandra},
   title = {Modeling negotiation subdialogues},
   booktitle = {30th Annual Meeting of the Association for Computational Linguistics},
   pages = {193-200},
   year = {1992},
   type = {Conference Proceedings}
}

@article{subd2,
   author = {Litman, Diane J and Allen, James F},
   title = {A plan recognition model for subdialogues in conversations},
   journal = {Cognitive science},
   volume = {11},
   number = {2},
   pages = {163-200},
   ISSN = {0364-0213},
   year = {1987},
   type = {Journal Article}
}

@article{pomdp,
   author = {Young, Steve and Gašić, Milica and Thomson, Blaise and Williams, Jason D},
   title = {Pomdp-based statistical spoken dialog systems: A review},
   journal = {Proceedings of the IEEE},
   volume = {101},
   number = {5},
   pages = {1160-1179},
   ISSN = {0018-9219},
   year = {2013},
   type = {Journal Article}
}

@article{pomdp1,
   author = {Williams, Jason D and Young, Steve},
   title = {Partially observable Markov decision processes for spoken dialog systems},
   journal = {Computer Speech \& Language},
   volume = {21},
   number = {2},
   pages = {393-422},
   ISSN = {0885-2308},
   year = {2007},
   type = {Journal Article}
}
\newpage
\appendix

\section{Taxonomy Codebook}
\label{app:taxanomony}
Table~\ref{tab:codebook_full} shows the full taxonomy codebook with detailed definitions and illustrative examples.

\begin{table*}[t]
\centering
\small
\sloppy
\caption{Full Taxonomy Codebook}\label{tab:codebook_full}
\begin{tabular*}{\textwidth}{@{\extracolsep{\fill}} p{0.22\textwidth} p{0.43\textwidth} p{0.27\textwidth} @{}}
\toprule
\textbf{Category} & \textbf{Definition} & \textbf{Representative Examples} \\
\midrule

\textbf{Deliberate Prioritisation} & The user has made a conscious judgment about which dimension of their decision deserves more or less reflective attention. The pattern reflects a reasoned assessment of relevance, importance, or usefulness — the user explains what they chose to focus on or not focus on, and why. & 
\parbox[t]{0.27\textwidth}{\raggedright
\textit{``Practicality is important when considering the welfare of a pet''} (ID\,280)\\[4pt] 
\textit{``I don't have past experiences with wanting a child, this is hypotheical and only about my feelings''} (ID\,67)} \\
\midrule

\textbf{Self-Diagnosed Focus} & The user has analytically identified a personal difficulty, tendency, or internal state that drives their reflective focus. They go beyond naming an emotion to explaining the mechanism — how a personal characteristic shapes their pattern. There is a diagnostic quality: the user is reasoning about their own psychology (System 2 reasoning about System 1 content). & 
\parbox[t]{0.27\textwidth}{\raggedright
\textit{``I focus a lot on feelings and how other people think because I worry about their perception of me''} (ID\,212)\\[4pt] 
\textit{``I feel like I have so much going on in my own head I sometimes can't take other peoples opinions on board.''} (ID\,128)} \\
\midrule

\textbf{Emotional Driver} & An emotion or unresolved emotional need explains the user's reflective focus. The emotion either pulls the user toward a dimension (e.g., sadness pulls toward feelings) or pushes them toward one (e.g., guilt pushes toward external obligations). The user names the emotion or emotional need as the reason without extensive analysis of how it operates (System 1 speaking). & 
\parbox[t]{0.27\textwidth}{\raggedright
\textit{``insecurity''} (ID\,167)\\[4pt] 
\textit{``Childhood trauma''} (ID\,191)} \\
\midrule

\textbf{Decision Nature} & The decision's inherent nature, the user's real-world circumstances, or their stage in the decision process explains the reflective focus. The pattern reflects what the decision IS or WHERE the user is in the decision process — not the user's psychology, emotions, or cognitive style.  & 
\parbox[t]{0.27\textwidth}{\raggedright
\textit{``out of necessity''} (ID\,135)\\[4pt] 
\textit{``You have to be practical in being able to live on the money you earn''} (ID\,246)} \\
\midrule

\textbf{Thinking Style} & The user's habitual way of approaching problems or decisions explains the reflective focus. The pattern reflects a stable personal characteristic — how they think generally — not something specific to this decision or their current emotional state. & 
\parbox[t]{0.27\textwidth}{\raggedright
\textit{``just the way I am''} (ID\,295)\\[4pt] 
\textit{``I am fact / risk driven in decision making''} (ID\,301)} \\
\midrule

\textbf{Interaction Attribution} & The user attributes the observed coverage pattern to the agent's questions, the conversational structure, the study design, or their expectations about the interaction format — rather than to their own reflective tendencies or the nature of their decision. &  
\parbox[t]{0.27\textwidth}{\raggedright
\textit{``The bot encouraged me to open up and think about my feelings in a constructive way.''} (ID\,282)\\[4pt] 
\textit{``it's just the way the conversation has flowed''} (ID\,206)} \\
\midrule

\textbf{No Accessible Reason} & The user recognises the observed pattern but cannot articulate why it exists. They express genuine uncertainty, possibly offering vague speculation without conviction. The reason for the pattern is not consciously available to the user. The user attempted to interpret but could not. & 
\parbox[t]{0.27\textwidth}{\raggedright
\textit{``not sure''} (ID\,126)\\[4pt] 
\textit{``Not really no''} (ID\,169)} \\
\midrule

\textbf{Content-Level Response} & The user did not attempt to interpret the pattern. Instead of explaining why their reflection has a particular distribution, they responded with decision content — narrating their situation, adding new information about their decision, or continuing to reflect on the decision topic. The metacognitive step of interpreting one's own reflective process did not occur. & 
\parbox[t]{0.27\textwidth}{\raggedright
\textit{``I have started a business before''} (ID\,308)\\[4pt] 
\textit{``I didn't mention that its a lot harder for me to conceive because of my health conditions, I have\ldots''} (ID\,227)} \\
\midrule

\textbf{Pattern Disputed} & The user rejects the agent's characterisation of their reflection. They either see a different pattern than what the agent described, believe the agent miscategorised their responses, or feel the observation doesn't accurately represent their thinking. & 
\parbox[t]{0.27\textwidth}{\raggedright
\textit{``I was focusing on experiences, not feelings''} (ID\,210)\\[4pt] 
\textit{``I do feel like I've reflected deeply from an emotional standpoint as well as a practical one''} (ID\,139)} \\
\bottomrule
\end{tabular*}
\end{table*}

\section{LLM Double-Coding Setup}

\subsection*{B.1 Prompt Template}

The prompt~\ref{fig:classification_prompt} was used for double-coding. Placeholders \texttt{\{observation\_text\}} and \texttt{\{interpretation\}} are filled with the agent observation and the participant's interpretation for each moment.

\begin{figure*}[t]
\begin{lstlisting}[
  basicstyle=\small\ttfamily,
  breaklines=true,
  frame=single,
  rulecolor=\color{gray!40},          % Makes the outer border line subtle
  backgroundcolor=\color{gray!10},    % Restores the grey box background
  xleftmargin=5pt,
  xrightmargin=5pt,
  framesep=8pt                        % Adds padding between the text and the grey border
]
Classify this user's interpretation into exactly one category.

Categories:
1. Deliberate Prioritisation --- User made a conscious judgment about where to focus. Example: "My feelings shouldn't outweigh what I can provide for the bird."
2. Self-Diagnosed Focus --- User analyses HOW a personal difficulty drives the pattern. Must contain causal reasoning (because, makes me, stems from, leads to). Example: "My anxiety has a massive role and it makes me focus on feelings."
3. Emotional Driver --- User names an emotion or emotional need as the reason, WITHOUT explaining how it operates. Example: "Guilt." / "I want to be happier."
4. Decision Nature --- The decision's situation or circumstances explain the focus, not the user's psychology. Example: "It's a financial decision." / "Social pressures are the driving factors."
5. Thinking Style --- A stable personal trait or habit explains the focus. Example: "That's how my brain works." / "I am an empathetic individual."
6. Interaction Attribution --- The agent's questions or the study design caused the pattern. Example: "The chatbot guided me there."
7. No Accessible Reason --- User tried to explain but couldn't. Example: "I'm not sure." / "Not really."
8. Content-Level Response --- User gave decision content instead of interpreting the pattern. Example: "I have started a business before."
9. Pattern Disputed --- User rejects the agent's observation as inaccurate. Example: "You misinterpreted my thoughts."

Key distinction: Self-Diagnosed Focus requires causal/mechanistic reasoning about oneself. Emotional Driver just names the emotion. If unsure between these two, check: does the interpretation explain HOW the emotion shapes the pattern, or just NAME the emotion?

Agent's observation: {observation_text}
User's interpretation: {interpretation}

Respond in exactly this format:
CATEGORY: [category name]
REASONING: [one sentence]
\end{lstlisting}
\caption{LLM classification prompt template for taxonomy categorization.}
\label{fig:classification_prompt}
\end{figure*}

\subsection*{B.2 Evaluation}
\label{app:irr}
Four models were evaluated:
\begin{itemize}\itemsep0pt
  \item Qwen 2.5 14B 
  \item Mistral Small
  \item Phi-4
  \item Gemma 2
\end{itemize}

Table~\ref{tab:irr_configs} reports inter-rater reliability summary, Cohen's $\kappa$.

\begin{table*}[!htb]
\centering\small
\caption{IRR $\kappa$ across all evaluated configurations.}\label{tab:irr_configs}
\begin{tabular*}{\textwidth}{@{\extracolsep{\fill}}lllll@{}}
\toprule
\textbf{Model A} & \textbf{Model B} & \textbf{Human--A $\kappa$} & \textbf{Human--B $\kappa$} & \textbf{A--B $\kappa$} \\
\midrule
Qwen 2.5 14B & Mistral Small & 0.46 & 0.50 & 0.54 \\
Phi-4 & Gemma 2 & 0.46 & 0.42 & 0.58 \\
\bottomrule
\end{tabular*}
\end{table*}

\section{Per-Category IRR}
\label{app:perC-irr}
Table \ref{tab:per_cat_agreement} shows agreement rates between the human coder and Mistral Small (the best-performing model, $\kappa=0.50$, $n=199$ double-coded moments). 
\emph{n} is the number of moments the human coder assigned to each category. Agreement \% is computed as the proportion of those moments where Mistral agreed. \emph{Most common confusion} reports the category Mistral most often assigned instead, with count in parentheses.

\begin{table*}[!htb]
\centering\small
\caption{Per-category human--Mistral agreement}\label{tab:per_cat_agreement}
\begin{tabular*}{\textwidth}{@{\extracolsep{\fill}}lrrl@{}}
\toprule
\textbf{Category} & \textbf{\textit{n}} & \textbf{Agree\,\%} & \textbf{Most common confusion} \\
\midrule
Content-Level Response & 1 & 0\% & Decision Nature ($n$=1) \\
Decision Nature & 22 & 18\% & Deliberate Prioritisation ($n$=12) \\
Deliberate Prioritisation & 61 & 74\% & Decision Nature ($n$=5) \\
Emotional Driver & 28 & 79\% & Deliberate Prioritisation / Self-Diagnosed Focus ($n$=3) \\
Interaction Attribution & 11 & 82\% & Decision Nature / Deliberate Prioritisation ($n$=1) \\
No Accessible Reason & 11 & 36\% & Pattern Disputed ($n$=4) \\
Pattern Disputed & 7 & 71\% & Deliberate Prioritisation / Interaction Attribution ($n$=1) \\
Self-Diagnosed Focus & 42 & 55\% & Deliberate Prioritisation ($n$=10) \\
Thinking Style & 16 & 44\% & Deliberate Prioritisation / Emotional Driver / Self-Diagnosed Focus ($n$=3) \\
\bottomrule
\end{tabular*}
\end{table*}

\section{Human–Model Confusion Matrix}
\label{app:confusion}
Table~\ref{tab:confusion_matrix} illustrates full $9\times9$ confusion matrix for human coder vs.\ Mistral Small on the $n=199$ double-coded sample. Rows represent the human (true) label; columns represent the Mistral (predicted) label.

\begin{table*}[t]
\centering\footnotesize
\caption{Confusion matrix: human coder (rows) vs.\ Mistral Small (columns). Diagonal cells (shaded) indicate agreement.}\label{tab:confusion_matrix}
\begin{tabular*}{\textwidth}{@{\extracolsep{\fill}}l rrrrrrrrr@{}}
\toprule
\textbf{Human\,$\backslash$\,Mistral} & \textbf{CL} & \textbf{DN} & \textbf{DP} & \textbf{ED} & \textbf{IA} & \textbf{NA} & \textbf{PD} & \textbf{SF} & \textbf{TS} \\
\midrule
CL & 0 & 1 & 0 & 0 & 0 & 0 & 0 & 0 & 0 \\
DN & 0 & \textbf{4} & 12 & 3 & 0 & 1 & 1 & 1 & 0 \\
DP & 0 & 5 & \textbf{45} & 4 & 4 & 1 & 2 & 0 & 0 \\
ED & 0 & 0 & 3 & \textbf{22} & 0 & 0 & 0 & 3 & 0 \\
IA & 0 & 1 & 1 & 0 & \textbf{9} & 0 & 0 & 0 & 0 \\
NA & 0 & 1 & 2 & 0 & 0 & \textbf{4} & 4 & 0 & 0 \\
PD & 0 & 0 & 1 & 0 & 1 & 0 & \textbf{5} & 0 & 0 \\
SF & 0 & 0 & 10 & 6 & 0 & 2 & 0 & \textbf{23} & 1 \\
TS & 0 & 0 & 3 & 3 & 0 & 0 & 0 & 3 & \textbf{7} \\
\bottomrule
\end{tabular*}
\vspace{4pt}
\begin{flushleft}\scriptsize
CL = Content-Level Response. \quad
DN = Decision Nature. \quad
DP = Deliberate Prioritisation. \quad
ED = Emotional Driver. \quad
IA = Interaction Attribution. \quad
NA = No Accessible Reason. \quad
PD = Pattern Disputed. \quad
SF = Self-Diagnosed Focus. \quad
TS = Thinking Style. \quad
\end{flushleft}
\end{table*}

\section{Taxonomy Distribution}
\label{app:taxDis}
Table~\ref{tab:taxonomy_dist} shows the distribution of the 232 coded pause moments across the nine taxonomy categories, with sub-breakdowns by moment position (opening vs.\ mid-session) and agent alignment.

\begin{table*}[!htb]
\small 
\centering
\caption{Full taxonomy distribution with position and alignment breakdown.}
\label{tab:taxonomy_dist}
\begin{tabular*}{\textwidth}{@{\extracolsep{\fill}}l rrrrrr@{}}
\toprule
\textbf{Category} & \textbf{\textit{n}} & \textbf{\%} & \textbf{Opening} & \textbf{Mid-session} & \textbf{Aligned} & \textbf{Misaligned} \\
\midrule
Deliberate Prioritisation & 64 & 27.6\% & 14 & 50 & 33 (52\%) & 31 (48\%) \\
Self-Diagnosed Focus      & 46 & 19.8\% & 18 & 28 & 19 (41\%) & 27 (59\%) \\
Emotional Driver          & 32 & 13.8\% &  7 & 25 & 12 (38\%) & 20 (62\%) \\
Decision Nature           & 26 & 11.2\% & 10 & 16 & 11 (42\%) & 15 (58\%) \\
Thinking Style            & 20 &  8.6\% &  3 & 17 & 11 (55\%) &  9 (45\%) \\
Interaction Attribution   & 15 &  6.5\% &  2 & 13 &  9 (60\%) &  6 (40\%) \\
No Accessible Reason      & 16 &  6.9\% &  2 & 14 &  6 (38\%) & 10 (62\%) \\
Content-Level Response    &  4 &  1.7\% &  3 &  1 &  2 (50\%) &  2 (50\%) \\
Pattern Disputed          &  9 &  3.9\% &  3 &  6 &  2 (22\%) &  7 (78\%) \\
\midrule
\textbf{Total}            & 232 & 100.0\% & 62 & 170 & 105 (45\%) & 127 (55\%) \\
\bottomrule
\end{tabular*}
\end{table*}


\section{Decision Topic Distribution}
\label{app:decisionTopics}
Distribution of the 13 decision topics represented in the sample, with counts of participants and pause moments, and the most frequent taxonomy category within each topic is shown in Table~\ref{tab:decision_topics}.

\begin{table*}[!htb]
\small 
\centering
\caption{Decision topics by participant count, moment count, and dominant taxonomy category.}
\label{tab:decision_topics}
\begin{tabular*}{\textwidth}{@{\extracolsep{\fill}}l r r l@{}}
\toprule
\textbf{Decision Topic} & \textbf{Participants} & \textbf{Moments} & \textbf{Most common category} \\
\midrule
Start a new job/position & 16 & 61 & Deliberate Prioritisation \\
Quit a job/position & 8 & 30 & Self-Diagnosed Focus \\
Start a new business & 6 & 25 & Deliberate Prioritisation \\
Get a pet & 5 & 20 & Deliberate Prioritisation \\
Move to a new city & 5 & 18 & Decision Nature \\
Have/adopt a child & 4 & 13 & Decision Nature \\
Pursue a degree & 4 & 16 & Deliberate Prioritisation \\
End a romantic relationship & 3 & 9 & Deliberate Prioritisation \\
Buy a home & 3 & 11 & Deliberate Prioritisation \\
Get married & 2 & 6 & Pattern Disputed \\
Move to a new country & 2 & 6 & Self-Diagnosed Focus \\
Care for a family member & 2 & 8 & Thinking Style \\
Begin a romantic relationship & 2 & 9 & Emotional Driver \\
\bottomrule
\end{tabular*}
\end{table*}


\end{document}